\documentclass[11pt]{article}

\usepackage{graphicx}

\title{Conductance Phases in Aharonov-Bohm Ring Quantum Dots}
\author{A. Yahalom$^a$ and R. Englman$^{a,b}$\\
 $^a$ College of Judea and Samaria, Ariel 44284, Israel\\
 $^b$ Department of Physics and Applied Mathematics,\\
 Soreq NRC,Yavne 81800,Israel\\
e-mail:  asya@yosh.ac.il; englman@vms.huji.ac.il;}

\begin{document}
\maketitle

\newcommand{\beq} {\begin{equation}}
\newcommand{\enq} {\end{equation}}
\newcommand{\ber} {\begin {eqnarray}}
\newcommand{\enr} {\end {eqnarray}}
\newcommand{\eq} {equation}
\newcommand{\eqs} {equations }
\newcommand{\mn}  {{\mu \nu}}
\newcommand{\sn}  {{\sigma \nu}}
\newcommand{\rhm}  {{\rho \mu}}
\newcommand {\SE} {Schr\"{o}dinger equation}
\newcommand{\sr}  {{\sigma \rho}}
\newcommand{\bh}  {{\bar h}}
\newcommand {\er}[1] {equation (\ref{#1}) }
\newcommand {\erb}[1] {equation (\ref{#1})}
\newcommand{\gb} {{\bf \gamma}}
\newcommand{\gcrb}  {{\bf \gamma^+}}
\newcommand{\gd} {{\dot \gamma}}
\newcommand{\gcr} {{\gamma^+}}
\newcommand{\gcrd} {{ \dot \gamma^+}}
\newcommand{\ro} {{ \gamma  \gamma^+}}
\newcommand{\gf} {{\gamma_1}}
\newcommand{\gs} {{\gamma_2}}

\begin{abstract}
The regimes of {\it growing} phases (for electron numbers
$N\approx0-8$) that pass into regions of {\it self-returning}
phases (for $N>8$), found recently in quantum dot conductances by
the Weizmann group are accounted for by an elementary Green
function formalism, appropriate to an equi-spaced ladder structure
(with at least \emph{three} rungs) of  electronic levels in the
quantum dot. The key features of the theory are {\it physically} a
dissipation rate that increases
 linearly with the level number (and tentatively linked to
 coupling to longitudinal optical phonons) and a set of Fano-like meta-stable levels,
 which disturb the unitarity, and  \emph{mathematically} the
change over of the position of the complex transmission
amplitude-zeros from the upper-half in the complex gap-voltage
plane to the lower half of that plane. The two regimes are
identified with (respectively) the Blaschke-term and the
Kramers-Kronig integral term in the theory of complex variables.
 \end{abstract}

 ~

 {\it PACS numbers: 03.65.Vf}

 ~

Keywords: Quantum dots, Aharonov-Bohm effect, Green function,
Hilbert Transform.

~

 \section{Two Phase Regimes}
Following a theoretical prediction in {\cite {GefenIA}, a
 pioneering experimental determination of the phase evolution
in quantum dots subject to the Aharonov-Bohm effect was made by
the Heiblum-led Weizmann group, as e.g. in \cite {YacobiHMS}-\cite
{JiHS}. The same group has recently come up with an interesting
development and  a physical description \cite{AvinunHZMU}, which
throw fresh light on their results (previous and recent). They
showed (cf. their figures 4-6) that as the gate voltage ($V_p$ in
their notation) increases and more electrons are entering the
quantum dot, the phase of the conductance evolves in the following
manner:

Initially, for a number of electrons N in the quantum dot up to 8
($N\leq8$) the phases corresponding to each N increase in a
stepwise fashion, following which, as $N>8$, the phases return
continuously to their original value (make a phase lapse).

Other manifestations of the absence of "phase rigidity" (meaning,
the discontinuous switch of phases between $0$ and $\pm \pi$ ,
connected to unitarity) in the phase-coherent transport across
quantum dots were observed in \cite{KobayashiAKI}.

On the theoretical front, the unexpected phase-behavior of the
experiments have resulted in numerous theoretical efforts, several
of which included investigation of the Kondo-effect (e.g.,
\cite{OregG}). Other works were directed at an analysis of the
results in terms of the Landauer-B\"uttiker formalism of
conductivity, which then led to consideration of the transmission
amplitude $t^{QD} (U)$ as function of the gap voltage U \cite
{HackenbroichW}-\cite{SilvestrovI}. The complicated geometry of
the experiments necessitated the inclusion in the theory of
several channels and the couplings between these \cite{EntinAK},
as well as a detailed analysis of the phase that was being
observed \cite{AharonyEHI}. A qualitative effect of changes in
both the transmission probability and the phase was theoretically
found when the signs were changed in some dot-lead coupling matrix
elements\cite {SilvaOG}. More recently, the ingoing-outcoming
coupling  asymmetry was studied more comprehensively, again in a
two level system \cite {GolosovG}.  A selective choice of the
experimental phase-conductance results obtained in \cite
{JiHSMS,JiHS} was matched with use of the Friedel sum rule in
\cite {JerezVL}, without accounting for the transition between the
phase-growth and the phase lapse regimes.

 Quantum dot-ring transmission is theoretically related to transmission
 in quantum wave-guides \cite{XuS,RyuC}. The latter was studied in
 \cite {Price}-\cite{ShaoPL}, which noted the existence of pole-zero
 pairs in the transmission amplitude (as function of the incident
 wave energy), and especially the changes that occurred with resonant attachments (stubs)
 to the wave guide.  An interesting finding was made in \cite
 {PorodSL}, that  in wave-guides some geometric changes (like attaching stubs) are formally
 equivalent to coupling between discrete and a continuum of
 states (the Fano-effect \cite{Fano}). For Aharonov-Bohm interferometric devices including quantum
  dots  the interrelation between geometry and Fano-states  was formulated in  \cite{{EntinAIL}}, and recent experiments
  were interpreted in terms of the Fano-effect \cite{KobayashiAKI}.
  We stress here this theoretical equivalence,
  since many previous explanations of the anomalous phase-behavior in quantum dots
concentrated on the geometrical aspect, whereas in the following
 theory the breakdown of unitarity is traced to decay of conducting levels and to the meta-stability
 electronic states lying above the quantum dot well. The essential modification that this paper
  makes in the previous treatments of the Fano-effect is thus that the higher-lying states are assumed to
  possess a short life time, so that the imaginary part dominates the energy denominator. (Under
   this assumption it makes no difference whether the higher lying states form a continuum or
    are discrete, as we propose for simplicity.) Establishing the causes of the meta-stability
    of the high lying states  and of the level-dependent decay rate in the lower lying states
    is not a primary aim of this work, which is largely phenomenological. Certain considerations indicate
    that both are due to coupling to longitudinal optical (LO) phonons and this
    speculative idea is described in section 5.

    Though apparently far removed from the physics of quantum dots, it turns out the the question
    of zeros and poles of the transmission amplitude or of a Greens function (the equivalence between
     which was demonstrated in \cite{ShaoPL}) plays an important role in the explanation of the phase behavior.
     This was heralded in several previous works (e.g., \cite {Deo}), but the present theory does this in
      a more comprehensive form, namely by use of Hilbert transform (in section 3) and through
      a compact representation of the transmission amplitude (in section 4) .

 Though the simple theory presented below is implemented by ad hoc
 assignment of some parameter values  and needs to
 be amplified to fill in several physical  details, it seems that it contains the
 answer to the question: What lies behind the strange phase
 behavior? A view expressed in \cite {AvinunHZMU} is that the so far available
 theories are short of providing an answer \cite {quote}. Earlier experimental
  data of \cite{SchusterBHMUS} for returning phases were rather
  precisely correlated with the observed conductivities by the
  present authors, using a parameter-free method which was the
  precursor of the present work \cite{EnglmanY}. In the
 Conclusion section of this work, we summarize the differences
 between the here proposed theory and those of other researchers.

 \section{Electron Transmission Amplitude}

The properties of the quantum dot (spin-less) electron
transmission function can be best understood in terms of the
theory of Hackenbroich and Weidenmuller \cite{HackenbroichW}. For
the sake of completeness we repeat here their end result.

\subsection{System Hamiltonian}

The system under consideration is composed of three sub-systems:
\begin{enumerate}
    \item The leads.
    \item The Aharanov-Bohm system not containing the dot.
    \item The quantum dot.
\end{enumerate}
The entire Hamiltonian of the system can be described by: \beq
H=H_0 + H_T \enq $H_0$ describes the totally disconnected system
and is given by: \beq H_0 = \sum_{akr} \epsilon_{ak}^{r}
c_{ak}^{r\dag} c_{ak}^{r} + \sum_{i} \epsilon_{i} d_{i}^{\dag}
d_{i} + \sum_{j} E_{j} q_{j}^{\dag} q_{j}+U_{ES} \enq $r$ denotes
the leads, $a$ runs over the channels in each lead, $k$ over the
longitudinal wave numbers, and $\epsilon_{ak}^{r}$  is the
corresponding energy. The energies of the single particle states
within the rings and within the dot are labelled by $\epsilon_{i}$
and $E_{j}$ respectively. $E_{j}$ is assumed to depend
parametrically on $U$. In the formal theory to follow, $U$ is a
complex quantity, whose real part $Re U$ is identified with the
experimentally manipulated plunger voltage $V_p$. $U_{ES}$ is the
electrostatic charging energy of the dot.

The coupling Hamiltonian $H_T$ has the form: \beq H_T =
\sum_{akir} W_{ai}^{r} (k) c_{ak}^{r\dag} d_i +  \sum_{ijp}
V_{ij}^{p} q_{i}^{\dag} d_j + H.C. \enq $W$ describes the coupling
between ring and leads, $V$ describe the much smaller coupling
between ring and dot.
 $p=L,R$ labels either side of the dot. In
actuality, we should allow for more exit channels than just the
two ($L$ and $R$) for the dot, corresponding to the experimental
arrangements in, e.g., \cite{AvinunHZMU}. We shall account for
these by including them in the postulated "high lying" energy
levels (see below in \er{td} and recall the discussion on the
equivalence between stub and Fano-states effects in our opening
section).

The transmission amplitude $t_{ab}(E)$ through the ring for an
electron entering the ring via channel $b$ in lead two, and
leaving it via channel $a$ in lead one is derived in
\cite{HackenbroichW}. We separate this as \beq t_{ab} =
t^0_{ab}+t^{QD}_{ab}\label{tab}\enq into the ring transmission and
the transmission $t^{QD}_{ab}$ across the quantum dot and treat
first the former.

\subsection{Aharonov-Bohm ring transmission}

The transmission matrix across the ring is expressed by \beq
t^0_{ab}= -2i\pi \sum_{ik} W^1_{ai} (D^0)^{-1}_{ik} W^{2*}_{bk}
\enq with the matrix $(D^0)_{ik}$ defined by: \beq (D^0)_{ik} = (E
- \epsilon_i) \delta_{ik} + i \pi \sum_{ct} W_{ci}^{t*} W_{ck}^{t}
\label{D0} \enq When the ring is feeded by the lead's reservoir
filled up to the Fermi energy $E_f$, one can replace $E$ in \er
{D0}by $E_f$. In the presence of a magnetic field threading the
circuit, the ring transmission amplitude will acquire an
Aharonov-Bohm phase factor.

\subsection {Quantum dot transmission}
We now turn our attention to the second term in \er {tab}. In the
case that repeated zig-zagging of carriers between the leads can
be ignored, this is the term whose magnitude and phase are
obtained in an Aharonov-Bohm interference measurement
\cite{AharonyEHI}. For simplicity, we drop the channel labels
$a,b$.

 We model the quantum dot as an electronic system having a ladder structure, i.e.
  $N_{el}$ equi-spaced level, interacting with some dissipative reservoir, say the LO
  phonons in the dot \cite{BeckelmannB}-\cite{StauberZC}. For quantum dots typified by those
   in the experiments discussed, the number of available levels is of the order
  of $100$ and their spacing is $40~\mu eV$ \cite{{Hackenbroich}}. We shall
  subdivide these levels
 into $N_{low lying}$ bound states, inside the quantum well and having an equi-spaced ladder
  structure, and a set of $N_{high lying}$ localized, meta-stable ("almost-bound")
  states, above the well \cite {Baym, HiroseMW}. The effect of these levels on the low lying
  levels is similar to the continua that feature in the Fano-effect. For
  simplicity, we take the number of these levels ($N_{high lying}$)
  finite.

  We next write the transmission
  amplitude across the dot within a wide band approximation,
 as described in \cite{HackenbroichW}. The limitations in applying
 the Hackenbroich-Weidenm\"uller approach to the experiments \cite {YacobiHMS}-
 \cite{AvinunHZMU} have been noted in \cite {Hackenbroich}
 (section 4.3.1). On the other hand, the observed regular peak  structures in
 some of these papers indicate that the following sum of
 Breit-Wigner terms should form an approximation to the
 transmission amplitude (at least, close to resonances).
  \beq
t^{QD}(U)= -i
 G\sum_{n=1}^{N_{el}}\frac{1}{E-E_f+U-<n|H_0|n>-R(E-E_f,n)}\label{t1}\enq
 $G$ is a single parameter
characterizing the scattering across the dot and is equivalent to
$2\pi W^2$ introduced above, (as before) $E_f$ is the Fermi energy
in the leads, $U$ is the gap voltage parameter in suitable units
whose real part $V_p$ is the experimentally manipulated depth of
the dot-well (however, we shall occasionally use $U$ also when we
mean its real part), $<n|H_0|n>=n$ is the electronic level energy
in suitable units, $R$ is the complex self-energy of the $n'$th
dot level, including also the coupling of the electrons to the
environment (erstwhile, the phonons and the stubs). Note that $U$
is {\it not} the Hubbard repulsion parameter, which will not be
explicitly taken into account, except for its presence in the self
energy $R$, which will also incorporate off-diagonal terms \cite
{SilvaOG}.

 For the self-energy $R=R'+iR"$ we now introduce our main
assumption that its imaginary part scales linearly for low lying
levels with the electronic level height \beq R"=-\gamma n
~~(0<\gamma\ll 1) \label{ImR} \enq For higher lying levels we
assume that the phonon electron coupling mechanism is so
efficient, that $|R''| >> |U-<n|H_0|n>-R|$. The width of these
levels is extremely large, so that the dependence of $U$ on the
contribution by those levels to $t^{QD}(U)$ is negligible. (This
is different from the usual treatment of the Fano-effect  in which
the contribution of the continuum is energy dependent
\cite{Fano}.)

 The $t^{QD}(U)$
terms can thus be dissected to two terms as follows: \beq
t^{QD}(U)= t_h^{QD} + t_l^{QD} (U)
 \label{t1b}
 \enq
in which we artificially disregard intermediate cases. In the
above equation:
 \beq
t_l^{QD}(U)= -i
G\sum_{n=1}^{N_{lowlying}}\frac{1}{E-E_f+U-<n|H_0|n>-R(E-E_f,n)}
 \label{tc}
 \enq
 and
\beq t_h^{QD}= G
\sum_{n=N_{lowlying}+1}^{N_{highlying}}\frac{1}{R''(E-E_f,n)}
 \label{td}
\enq
 We next use expression \er{t1b} to calculate $t^{QD}(U)$, the
quantum dot
 transmission coefficient as function of the
the gap voltage $U$. Figures 1 and 2 show the results, with the
following choice of parameters (having put $E=E_f$): \beq
N_{lowlying}=34,~\frac{t_h^{QD}}{G}=1.35,~\gamma =.0086,~R'=-8.5~
\label{param} \enq The figures show clearly the peaked structure
of the absolute value of the transmission amplitude (the
visibility or scaled $\mid$ conductance $\mid$) at subsequent
electron fillings and the radical change of character in the
phase-behavior. Due to our chosen fitting of the energy shift
parameter ($-8.5$) and of $\gamma=0.0086$ in \er {param}, this
change occurs just at the experimental value of \cite
{AvinunHZMU}.
\begin{figure}
\includegraphics{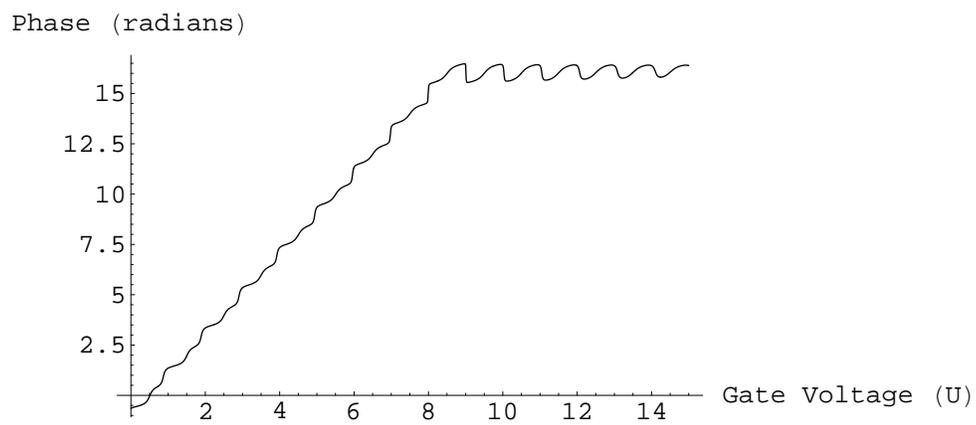}\\
\caption{The phase of the transmission amplitude for the parameters given by \er{param}}
\label{Phase}
\end{figure}
\begin{figure}
\includegraphics{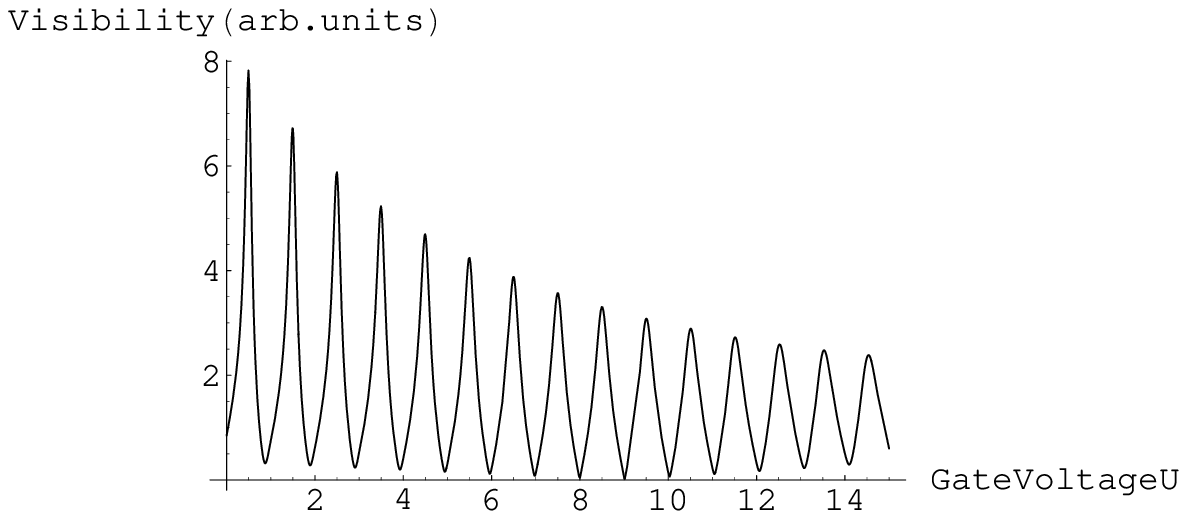}\\
\caption{The absolute value of the transmission amplitude (in
arbitrary units) for the parameters given by \er{param}}
\label{Vis}
\end{figure}

\subsection{Numerical properties of the transmission amplitude}

Inserting the numerical parameters from \er{param}into \er{t1b} we
obtain the quantum dot transmission amplitude as \beq
t^{QD}(U)/G\approx 1.35
-i\sum_{n=1}^{34}\frac{1}{U+8.5-(1-0.0086i)n}\label{expl}\enq The
poles (resonances) occur at such half-integral values of $U$, that
$U+8.5$  matches one of the integers $n$ in the denominator. The
widths increase with $n$. Across each resonance the phase
increases by $\pi$, as usual across  Breit-Wigner resonances.
However, unlike the latter, there are now also (complex) zeros,
passing which either add or subtract $\pi$ to the phase, depending
on whether the zero lies in the upper or the lower half $U$-plane.
The leading term ($1.35$ in \er{expl}), whose source is the higher
lying states, is essential for the existence of the zeros. It
turns out that to obtain the zeros around any $U$ (or $n-8.5$), it
is necessary to consider two more terms in the sum, one on each
side of the resonance. (One neighboring term is insufficient;
three or more terms are qualitatively unnecessary. This numerical
aspect distinguishes the present approach from several previous
ones, e.g. \cite {SilvaOG}, which considered only two resonances.
Some exceptions are \cite{AharonyEHI} and \cite {BaltinG}, which
however do not include the fast decaying levels.) It then emerges
that, for $U<8.5$ one finds three zeros in the upper half
$U$-plane which make up a total $6\pi$ increase over three
resonances; whereas, for $U>8.5$ one finds just one zero in the
lower half $U$-plane, which leads to a total of $2\pi$ phase
change. Merging {\it all} the resonances yields the curves shown
in figures $1$ and $2$.

\section {The General Significance of Complex-Zeros}

We now describe the formal basis of the above result, showing that
the change of behavior is not accidental, but rather required by
simple mathematical properties of the transmission amplitude
$t^{QD} (U)$ regarded as a function of the variable $U$:

 The underlying reason is that
just such behavior of phases is expected for a quantity $t^{QD} (U)$
that has the following properties (in addition to $t^{QD} (U)$
satisfying certain formal, analytical properties \cite {PaleyW,
Titchmarsh}):

$t^{QD} (U)$ has zeros in the upper half of the complex $U$-plane
for $Re U\leq8$ and has zeros in the lower half of the $U$-plane
for $Re U>8$. [As before, we have identified the real part of U
with a scaled gate-voltage $V_p$. The gate-voltage $V_p$ increases
the number $n$ of bound electrons in the quantum dot.]

Why is this so straightforward?

As shown immediately below, the phase evolution can be expressed
as a sum of (essentially) two terms: an integral term and the (so
called) Blaschke terms. The former shows structure (wiggles) of
phase return, but no net gain (i.e. it returns to the starting
value) and the latter shows net gains, phase growth (and no
structure). Precisely, the Blaschke terms arise from singularities
of $ \ln t^{QD} (U)$ in the upper-half-plane and the structure in
the integral comes from singularities of $ \ln t^{QD} (U)$ in the
lower half plane (due to continuity). Furthermore, both the
wiggles and the gain (in the phase) are tied to maxima in the
visibility ($|t^{QD} (U)|$), as in the experiments.

Thus the minimal property required of $t^{QD} (U)$ is that its complex zeros
lie in the upper half plane for $Re (U)$ less than $8$ and in the lower
half plane for $Re (U)$ larger than $8$. In the sequence we shall build up
at least one simple function $t^{QD} (U)$ that
has these properties, but there are obviously others, too.
\begin{figure}
\includegraphics{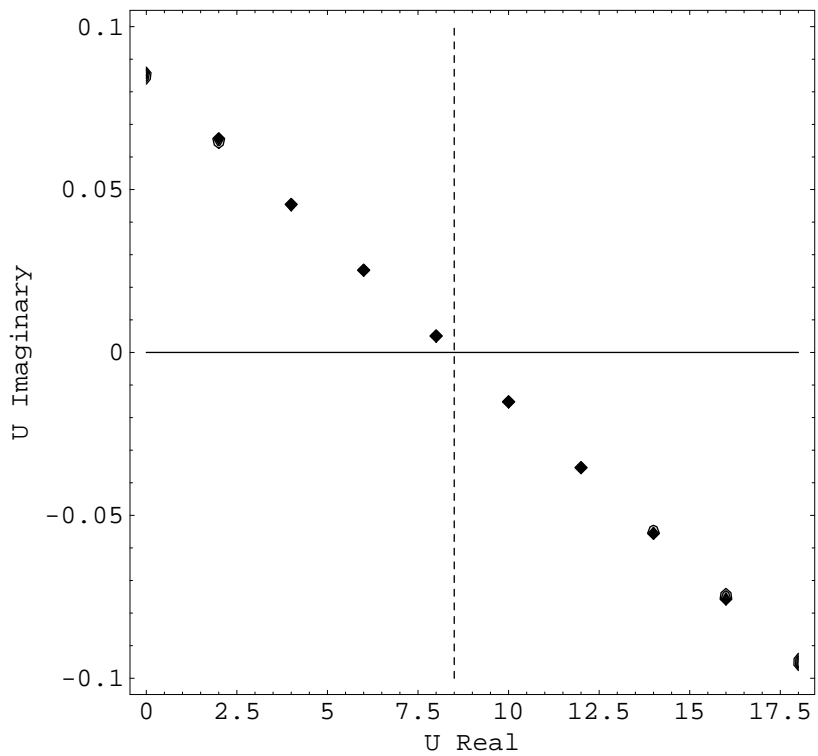}\\
\caption{Zeros of $t^{\infty}(U)$ (see \er{expansion} below) with parameter values
$ A=1,~B=2.5,~U_c=8.5,~\gamma=.01$.}
\label{zeros}
\end{figure}

\subsection{The Blaschke terms}

 Let us explain the "Blaschke-terms".  These arise if the well-known Kramers-Kronig
(KK) relations are applied to the {\it logarithm} of a regular
function $t^{QD} (U)$ of its argument $U$, rather than to $t^{QD}
(U)$ itself, as is usual. Then the zeros of $t^{QD} (U)$ add
singularities to the KK integrand and these have to be subtracted
in a manner that does not affect adversely the conditions that are
the basis of the KK relations. As a consequence (for real values
of U) one can express the argument (phase) of this function as
\beq arg~t^{QD}
(U)=-\frac{1}{\pi}P\int^{\infty}_{-\infty}\frac{dU'}{U'-U}
ln|t^{QD}(U')| + \Phi_B(U) \label{phase1} \enq where $P$
represents the principal part of the singular integral and
$\Phi_B(U)$ is the Blaschke-phase given as the sum of terms \beq
\Phi_B(U)= -i\sum_j ln \frac{U-U^j}{U-U^{j*}}\label{Blaschke}\enq
\cite{vanKampen}-\cite{MandelW}. Here $U^j$ are those zeros of
$t^{QD} (U)$ that lie in the upper half  of the complex $U$-plane
and $U^{j*}$ are their complex conjugates. (Actually,
\er{Blaschke} can be generalized for poles and branch-points in
the upper half-plane by suitably attaching (negative and
fractional) weights to each term in the sum, but since we shall
find that there are no poles or branch-points in the transmission
amplitude for the range of interest, we can disregard these
possibilities.)

Now if we look at the integral term, we see that it tends to $0$
for both $U\to -\infty$ and $U\to \infty$ (provided the
log-function has no singularities {\it on} the real $U$-axis).
Therefore, as claimed, this term cannot cause a net gain of the
phase, only some structure. Such structure will indeed occur
 when $|t^{QD} (U)|$ becomes small at some value of
$U$. It will have the form of a very sharp peak whenever a zero of
$t^{QD} (U)$ will be very close to the real axis. This will occur
when $\gamma$ is very small compared to the level spacing, as in
\er{param}, for which the level spacing was unity.

A different story are the terms in the Blaschke-phase. Each term
 will cause a step of $2\pi$ in the phase.

In the enclosed  drawings we show $|t^{QD} (U)|$ and
$\frac{1}{\pi}arg~ t^{QD} (U)$ both {\it vs} (the real part of)
$U$ on the same graph. In the visibility $|t^{QD} (U)|$ one sees
the peak structure  and in the phase: the initial steps (up to
$U=8$), followed (for $U$ above this value) by the rise and lapse
of the phases.

\section {A Compact Form of the Transmission Amplitude}
We now rewrite the preceding expression for the total transmission matrix $t^{QD} (U)$ (making
only an
approximation that will turn out to  have almost no effect on the results) and obtain a
compact, closed expression. From this we can deduce the relevant analytic
properties of $t^{QD} (U)$ almost by inspection.

Because we expect that for a given value of $U$ only a few (nearly resonant) terms in \er{tc}
will contribute, we extend the sum in \er{t1b} to $-\infty$ and $\infty$.
The resulting series can be summed to take a simple form
  \beq
 \frac{ t^{\infty}(U)}{t^{\infty}(U\rightarrow -\infty)}= \frac{1+A e^{-2\pi i (U-U_c)/(1 -i\gamma)}}{1 -B~e^{-2\pi i
(U-U_c)/(1 -i\gamma)}}
\label{C}
\enq
The algebra  is based on the result \cite{MacRobert}
\beq
\frac{1}{e^z-1}=-\frac{1}{2}+\sum_{n=-\infty}^{\infty}\frac{1}{z-2n\pi i}
\label{series}
\enq
from which follows the expansion of $t^{\infty}(U)$ as the series:
\ber
&\frac{ t^{\infty}(U)}{t^{\infty}(U\rightarrow -\infty)}  =
 \frac{1}{2}(1-\frac{A}{B}) -i(1+\frac{A}{B})\frac{(1-i\gamma)}{2\pi}&
\nonumber \\
&\cdot
\sum_{n=-\infty}^{\infty} \frac{1}{U-n-(U_c+\frac{\gamma}{2\pi}\ln~B)+i(\gamma~n+\frac{1}{2\pi}\ln~B)}&
\label{expansion}
\enr
Recalling now from \er{t1b} $ t^{QD}(U)= t_h^{QD} + t_l^{QD} (U)$, and noting the expression
for $t_l^{QD}$ in \er{tc}, we can make the following replacements:
\ber
& t_h^{QD}=\frac{t^{\infty}(U\rightarrow -\infty)}{2}(1-\frac{A}{B}), \qquad
2\pi G \simeq t^{\infty}(U\rightarrow -\infty)(1+\frac{A}{B})&
\nonumber \\
&R'=-U_c- \frac{\gamma}{\pi}\ln~B,\qquad
R"=-\gamma~n-\frac{1}{2\pi}\ln~B& \label{identif} \enr
 (In the second equation we have neglected the small and unimportant
quantity $-i\gamma$ before the sum.) Equation (\ref{identif}) will
lead to the following proportion between $A$ and $B$: \beq
\frac{A}{B}=\frac{\pi G -t_h^{QD}}{\pi G +t_h^{QD}}\label{AperB}
\enq

 The following values of the four parameters ($A,B,U_c,\gamma$) in the
function $t^{QD} (U)$ are compatible with the choices of the
parameter in \er{param}. \beq A=1,~B=2.5,~U_c=8.5,~\gamma=.0086
\label{param2} \enq
 The plotted
$t^{\infty}(U)$ with these parameters is shown in Fig. 4. The
result is virtually identical with that obtained for $t^{QD} (U)$
from the restricted sum in \er{t1}, in the gap voltage range of
figures 1 and 2. As already noted, the reason is that the
contributions to the infinite sum outside the restricted range are
negligible. The signal advantage of the compact form in \er{C},
over the partial sum in \er{t1}, is that the zeros and poles of
the transmission amplitude can be derived from the former
considerably simpler. We now obtain these zeros and poles, with
the parameters chosen in \er {param2}.
\subsection {Analysis of zeros and poles}
1)~ Zeros of \er{C}: These occur when the second term in the
numerator is  -1, so that \ber U & = &  U_c +
n+\frac{1}{2}-\frac{\gamma lnA}{2\pi}
-i\gamma[(n+\frac{1}{2})+\frac{ln A}{2\gamma \pi}]~~ (n=0~or~ a~signed~integer)\nonumber\\
& = & 8.5 + (n+\frac{1}{2})-0.0086i(n+\frac{1}{2})\nonumber \\  &
= & V_p(n) +0.0086i[8.5-V_p(n)]\label{zeros2}\enr where in the
second line we have inserted the parameter values from \er{param2}
and in the third line we have written $V_p(n)$ for the real part
of the $n$'th zero. For a small value of the decay rate $\gamma$
this will be the value of the observed gap voltage at the position
of the minimum. It is now apparent that for minima at gap voltages
below $8.5$ the zeros will be at positive imaginary parts of $U$,
while for gap voltages above $8.5$ the imaginary part will be
negative. (This was shown in Fig. 3.)

2)~Poles of \er{C}: For these the second term in the denominator
must be 1, giving \ber U & = &  U_c + m-\frac{\gamma
lnB}{2\pi} -i\gamma( m+\frac{ln B}{2\gamma\pi})~(m=0~or~ a~signed~integer)\nonumber\\
& = & 8.5 + m-0.0086i(m +17)\nonumber \\  & = & W_p(m)
-0.0086i[W_p(m)+8.5]\label{poles2}\enr where again (in the second
line) we have substituted the parameters and then have rewritten
the equation in terms of the observational gap voltages $W_p(n)$
at the maxima. (The small quantity $\frac{\gamma ln
B}{2\pi}\approx 0.001$ has been neglected.) It is now clear that
the poles lie in the lower half of the $U$-plane for all gap
voltages above $-8.5$. Gap voltages below this value are outside
the range of interest for the discussion of the experiments.

\subsection {Deductions from the compact form}
 The essential features of this form are
that {\it for values of the gate voltage $U$ that are
experimentally measured}

(1) there are no singularities (i.e.,
denominator zeros) in the upper complex-$U$ half plane, and

(2) for $Re U>U_c$ the zeros of $t^{QD} (U)$ are only in the
lower-half  of the complex $U$-plane (this is the phase-lapse
regime, identified with the integral part in the analytical
expression for the phase), whereas for $Re U<U_c$  there are zeros
in the upper-half of the complex $U$-plane (this is the
increasing-phase regime, identified with the Blaschke phase
terms). Important  in \er{C} are the parameter $U_c(=8.5)$ and
that $B(=2.5)>A (=1)$. The latter requirement removes the poles
from the wrong half-plane and,  by \er{AperB}, translates
immediately to the physical requirement that the high lying
(Fano-like) states' transmission amplitude $t^{QD}_h$ is
\emph{real} and positive.

Contrasting to our zeros, which are complex, the zeros  that were
found both in \cite {ShaoPL} and in \cite {SilvestrovI} were real.
In the last work it was indeed pointed out (on p. 106602-4) that
the reality was due to the time reversal invariance  of the
Hamiltonian , tied to an infinitely sharp $\pi$ phase jump,
whereas a finite width phase jump could be achieved by inter-level
thermal excitation. Alternatively, it could be obtained with a
nonzero $\gamma$ due to inelastic electron-phonon interactions,
which is the possibility envisaged here, and $B/A>1$ and which, by
\er{AperB}, is contingent to the virtual excitation to meta-stable
states.

(At this stage one may want to compare the functions of Table II
and the figures in the earlier paper \cite {EnglmanY} by the
present authors, in which the decay parameter $\gamma$ was $0$. A
more significant difference is that in the functions of \cite
{EnglmanY} the assumed regions of analyticity were the opposite to
that in the present article. The former choice is the natural one
if $U$ is identified with a "time-like" variable, whereas the
present choice is the proper one if $U$ is energy or frequency
like.)
\begin{figure}
\includegraphics{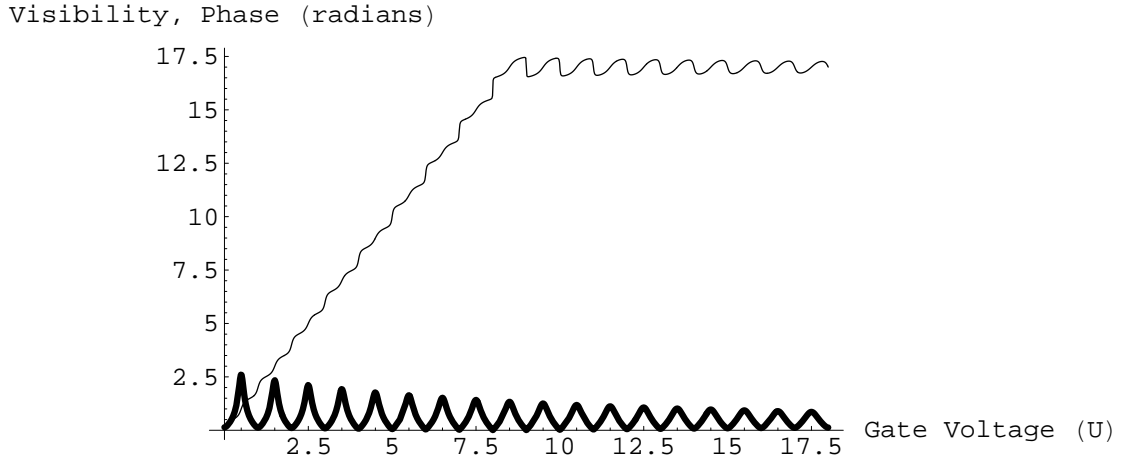}\\
\caption{Visibility amplitude ratio (thick line, arbitrary units)
and phase (thin line in radians) as function of gap voltage $Re
U$. $|t^{\infty}(U)|$ and arg$~t^{\infty}(U)$ are plotted from
\er{C} with parameter values $
A=1,~B=2.5,~U_c=8.5,~\gamma=.0086$.} \label{VisPha}
\end{figure}

Changes in some parameters can alter, e.g., the relative
magnitudes of the peaks. The slope of the phase-lapse in the figure is
proportional to the height of the minima in the visibility above
the origin. (This property was first predicted in \cite{EnglmanY}
and rediscovered in several subsequent papers.)

\subsection {A remark on the phase-step magnitude} The  phase shown in the drawing for the initial
(step-up) regime is not the same as in e.g. \cite{SchusterBHMUS}
or \cite{AvinunHZMU}, in that we predict a net phase gain of
$2\pi$ per peak, whereas the experimental phase steps seem  (in
most cases) to be less than this. If the discrepancy really
exists, the present interpretation may have to be withdrawn or be
changed in a way not clear to us just now. However, it seems that
the experimental phases are not traced quite precisely throughout
the step. Thus, when the visibility is near zero,  the phase
changes may not be properly recorded, but rather sawn together in
a continuous fashion so that part of the rise is lost.

\section {Speculating on the Decay Mechanism: Electron-LO Phonon Coupling ?}

Assuming a ladder-like structure for the low lying electronic
levels in the quantum dot, with level separation of unity (when
expressed in the units of $U$), our expression for $t^{\infty}(U)$
with $A=0$ can be simply understood as the  Green function of
broadened regularly spaced electronic states. The pre-exponential
factor $B$ and the $i\gamma$ part in the exponent then represent
the broadening of low lying levels. Were the former 1 and the
latter 0, we would have the Green function for a series of
equidistant, infinitely sharp electronic levels. However, our main
interest is in the zeros of the numerator. These arise because
$A\neq0$.

The phonon-bottleneck or its absence has long been under
consideration for the mechanism of decay of discrete electronic
levels in quantum dots \cite{BeckelmannB}-\cite {StauberZC}. It is
generally supposed that LO phonons in the dot of energy $\hbar
\omega_{LO}$ couple to the levels. It has also been noted that
when the electron level structure at some rung in the ladder gets
into near coincidence with the phonon energy, then a Rabi
splitting takes place. The physical meaning of this is that the
near-coincidental excited electron-level gets strongly admixed
with the ground electronic level in which one LO phonon is
excited. As a result, two admixture levels are formed, which are
separated by roughly the coupling energy between the electron and
the LO phonon. The condition for coincidence to occur at the $n_R$
(R for Rabi) electronic level is that \beq
 n_R \Delta\approx\hbar \omega_{LO}
 \label {coinc}
 \enq
where $\Delta$ is the  electronic energy separation.

We speculate that the decay in  $t^{QD} (U)$ reflects this
resonance condition i.e. $ \hbar \omega_{LO} = n_R \Delta\approx
N_{low lying}\Delta$, since above the low lying levels commences
the LO -phonon decay mechanism. We have not calculated the
transmission matrix of the coupled electron-LO optical excitation
(constituting a polaron), along the lines of
\cite{GlazmanS}-\cite{MachnikowskiJ}.
 With the estimates of \cite{Hackenbroich} that $\Delta = 40\mu eV $ (which may be a minimal
 estimate) and that there are
 $N_{lowlying}\approx 200$ electronic states up to the brim of the quantum dot
 well, one obtains $8 meV$ for the height of electronic levels, at
 which the phonon coupling causes an effective admixture. This is
 about a quarter (of $ \hbar \omega_{LO} = 36 meV$) where we would expect
 the electron-phonon coupling to be felt in GaAs \cite{StauberZC}.
 Our computations have stopped at $N_{lowlying}=34$, since
 the experimental range of scanned levels is considerably below this. No
 observable difference would be felt by extending the sum to
 $N_{lowlying}$ placed in the hundreds.

\section{Conclusion}

The two distinct regimes in the electron-transmission phase of an
Aharonov-Bohm arrangement containing a quantum dot, already
 present in  earlier experiments in \cite
{YacobiHMS}-\cite {JiHS} but recently definitively established in
\cite {AvinunHZMU}, have been explained by a model based on a
ladder of electronic levels  with increasingly faster decay from
higher levels up to a meta-stable continuum (or bunch of Fano-type
levels) with very short life-times. The decay mechanism is
tentatively surmised as due to LO phonons in the dot. Though a
Hamiltonian is postulated, its implementation in the transmission
amplitude is phenomenological.

Among the main new features of this work, not present in several
previous theories, are  the postulate of the large widths of
Fano-type states that are above the quantum well, the treatment of
geometrical effects (side-arms in the ring) on the same footing as
the admixture with meta-stable states to disrupt the unitarity
\cite{ShaoPL}, and the finding that at least three particle states
are needed to reproduce the observed phase behavior. In formal
terms, the two regimes of phases, increasing across the resonance
and those returning to former values are identified with zeros
(but not the poles!) of the complex transmission lying
(respectively) in the lower and upper half planes of the complex
energy (or gap voltage) variable. The absence of poles is
connected to the meta-stable state, but the zeros do not arise
from the usual Fano-form or from cancellation between adjacent
resonances.
\\
\\
\noindent
{\LARGE \bf Acknowledgements}
\\
\\
\noindent
 We thank Amnon Aharony, Yuval Gefen and Yuval Oreg for valuable remarks on
 our work and  the first for suggestions.

\end {document}